\documentclass[prl,twocolumn,showpacs]{revtex4-1}

\newcommand{\Eq}[1]{Eq.~(\ref{eq:#1})}

\newcommand{\Fig}[1]{Fig.~\ref{fig:#1}}
\newcommand{\Figs}[1]{Figs.~\ref{fig:#1}}

\newcommand{\expt}[1]{\left<#1\right>}
\newcommand{\f}{\mathbf{f}}
\newcommand{\gdot}{\dot\gamma}
\newcommand{\phieff}{\phi_\mathrm{eff}}
\renewcommand{\O}{{\cal O}}
\newcommand{\rr}{\mathbf{r}}

\newcommand{\tv}{\tilde \mathbf{v}}
\newcommand{\tvc}{\tilde v_c}
\newcommand{\tva}{\tilde v_a}
\newcommand{\tvs}{\tilde v_s}
\newcommand{\tvy}{\tilde v_y}
\newcommand{\tvcr}{\tilde v_\mathrm{cr}}
\newcommand{\tvone}{\tilde v_1}

\newcommand{\tvrms}{\tilde v_\mathrm{rms}}
\newcommand{\tvhalf}{\tilde v_{50}}
\renewcommand{\v}{\mathbf{v}}
\newcommand{\vtot}{\mathbf{v}^\mathrm{tot}}

\usepackage{graphicx}

\begin{document}
\title{Dissipation and velocity distribution at the shear-driven jamming transition}

\author{Peter Olsson}
\affiliation{Department of Physics, Ume\aa\ University, 901 87 Ume\aa, Sweden}

\date{\today}   
\begin{abstract}
  We investigate energy dissipation and the distribution of particle velocities at the
  jamming transition for overdamped shear-driven frictionless disks in two dimensions at
  zero temperature. We find that the dissipation is caused by the fastest particles and
  that the fraction of particles responsible for the dissipation decreases towards zero as
  jamming is approached. These particles belong to an algebraic tail of the velocity
  distribution that approaches $\sim v^{-3}$ as jamming is approached. We further find
  that different measures of the velocity diverge differently, which means that concepts
  like ``typical velocity'' may no longer be used---a finding that should have
  implications for analytical approaches to shear-driven jamming.
\end{abstract}

\maketitle

The hypothesis that the slowing down of the dynamics in systems as different as
supercooled liquids, granular materials, colloids, foams, and emulsions, have a common
origin in the properties of a critical point, point~J\cite{Liu_Nagel}, has inspired a
great amount of work on jamming the last decade. Several models have been used to try and
pinpoint the properties of this jamming transition. Some of them have centered around a
greatly simplified numerical model of spheres with contact-only interaction. One important
branch has been to examine the properties of randomly generated static
packings\cite{OHern_Silbert_Liu_Nagel:2003} whereas another has been to study the jamming
transition through simulations of elastic particles under steady shear\cite{Durian:1995}.

A key feature of jamming is the approach of the contact number $z$ to the isostatic number
$z_\mathrm{iso}$ which is just enough for mechanical stability. It has recently been
shown\cite{Lerner-PNAS:2012} that this is directly linked to the divergence of
$\eta_p\equiv p/\gdot$---the pressure equivalent of the shear viscosity. A related
phenomenon is the increase in particle velocity as
$\phi\to\phi_J$\cite{Heussinger_Berthier_Barrat:2010,Andreotti:2012}. This is related to
the distribution of particle displacements due to a small shear increment which has been
determined both in experiments of sheared granular materials\cite{Marty:2005} and in
quasistatic simulations\cite{Marty:2005, Heussinger_Berthier_Barrat:2010}. It was there
found that this distribution is sufficiently wide that the non-Gaussian parameter
$\expt{\Delta y^4}/3\expt{(\Delta y)^2}-1$ diverges as $\phi_J$ is approached from below,
granted that the shear step is sufficiently small.

In this Letter we show that there is more to the particle velocity distribution than has
so far been realized. Dissipation is mainly caused by the fastest particles and we find
that the fraction of particles that are responsible for the dissipation decreases towards
zero as jamming is approached. This behavior is related to an algebraic tail, $P(v)\sim
v^{-3}$, in the velocity distribution and we show that the velocity histograms determined
at the jamming density approach this limiting behavior as $\gdot\to0$. Since rheology and
dissipation are linked through power balance, the understanding of this phenomenon is
right at the center of the phenomenon of shear-driven jamming. We also note that this
finding has profound consequence for analytical approaches to jamming since it implies
that different measures of the velocity behave differently and that concepts like
``typical velocity'' therefore become useless.

Following O'Hern \emph{et al.}\cite{OHern_Silbert_Liu_Nagel:2003} we use a simple model of
bi-disperse frictionless soft disks in two dimensions with equal numbers of disks with two
different radii in the ratio 1.4. Length is measured in units of the diameter of the small
particles, $d_s$. We use Lees-Edwards boundary conditions\cite{Evans_Morriss} to introduce
a time-dependent shear strain $\gamma = t\gdot$. With periodic boundary conditions on the
coordinates $x_i$ and $y_i$ in an $L\times L$ system, the position of particle $i$ in a
box with strain $\gamma$ is defined as $\rr_i = (x_i+\gamma y_i, y_i)$.  The ordinary
velocity is $\vtot_i=\dot\rr_i$, but in the following we consider the non-affine velocity,
$\v_i = \vtot_i- \v_\mathrm{R}(\rr_i)$ where $\v_\mathrm{R}(\rr_i)\equiv\gdot y_i\hat x$
is a uniform shear velocity. With $r_{ij}$ the distance between the centers of two
particles and $d_{ij}$ the sum of their radii, the relative overlap is $\delta_{ij} = 1 -
r_{ij}/d_{ij}$ and the interaction between overlapping particles is $V(r_{ij}) = \epsilon
\delta_{ij}^2/2$; we take $\epsilon=1$. The force on particle $i$ from particle $j$ is
$\f^\mathrm{el}_{ij} = -\nabla_i V(r_{ij})$. The simulations are performed at zero
temperature.

We consider two different models for the energy dissipation. In both cases the interaction
force is $\f^\mathrm{el}_i = \sum_j \f^\mathrm{el}_{ij}$ where the sum extends over all
particles $j$ in contact with $i$, and the equation of motion is
\begin{equation}
  \f^\mathrm{el}_i +\f^\mathrm{dis}_i = m_i\ddot\rr_i.
\end{equation}
Most of our simulations have been done with the RD$_0$ (reservoir dissipation) model with
the dissipating force
\begin{equation}
  \f^\mathrm{dis}_{\mathrm{RD},i} = -k_d \v_i.
\end{equation}
We take $k_d=1$, $m_i=0$, and the time unit $\tau_0 = d_s k_d/\epsilon$.  We simulate
$N=65536$ particles with shear rates down to $\gdot=10^{-10}$. Checking for finite size
effects at $\gdot=10^{-9}$ we found no difference when using instead $N=262144$. The
equations of motion were integrated with the Heuns method with time step $\Delta
t=0.2\tau_0$.

Some additional simulations have also been done with the CD$_0$ model (CD for ``contact
dissipation'') with dissipation due to velocity differences of disks in
contact\cite{Durian:1995, Tighe_WRvSvH}. Details of these simulations may be found
elsewhere\cite{Vagberg_OT:jam-cdrd}.

\begin{figure}
  \includegraphics[bb=40 324 532 654, width=7cm]{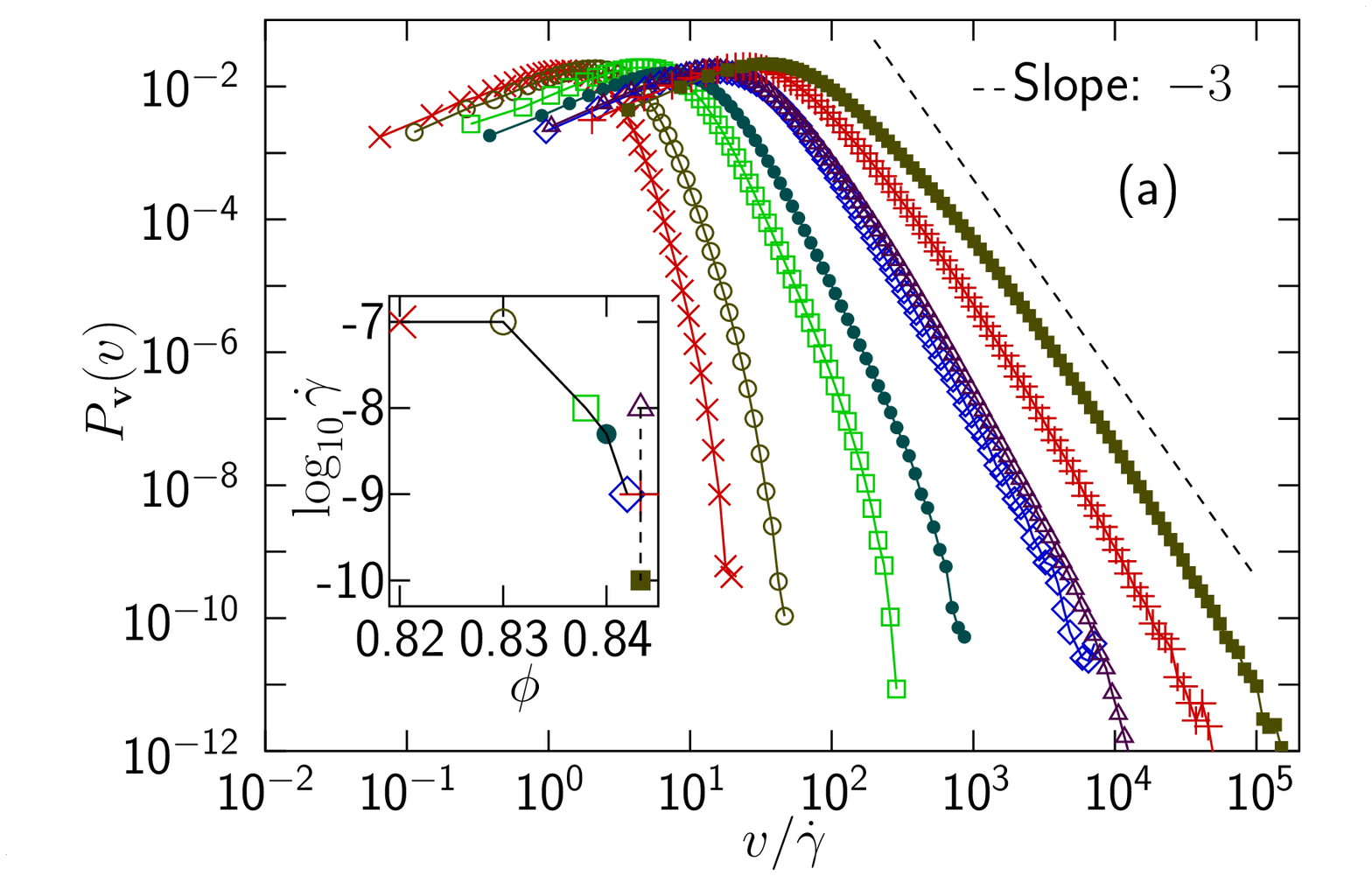}
  \includegraphics[bb=40 324 532 654, width=7cm]{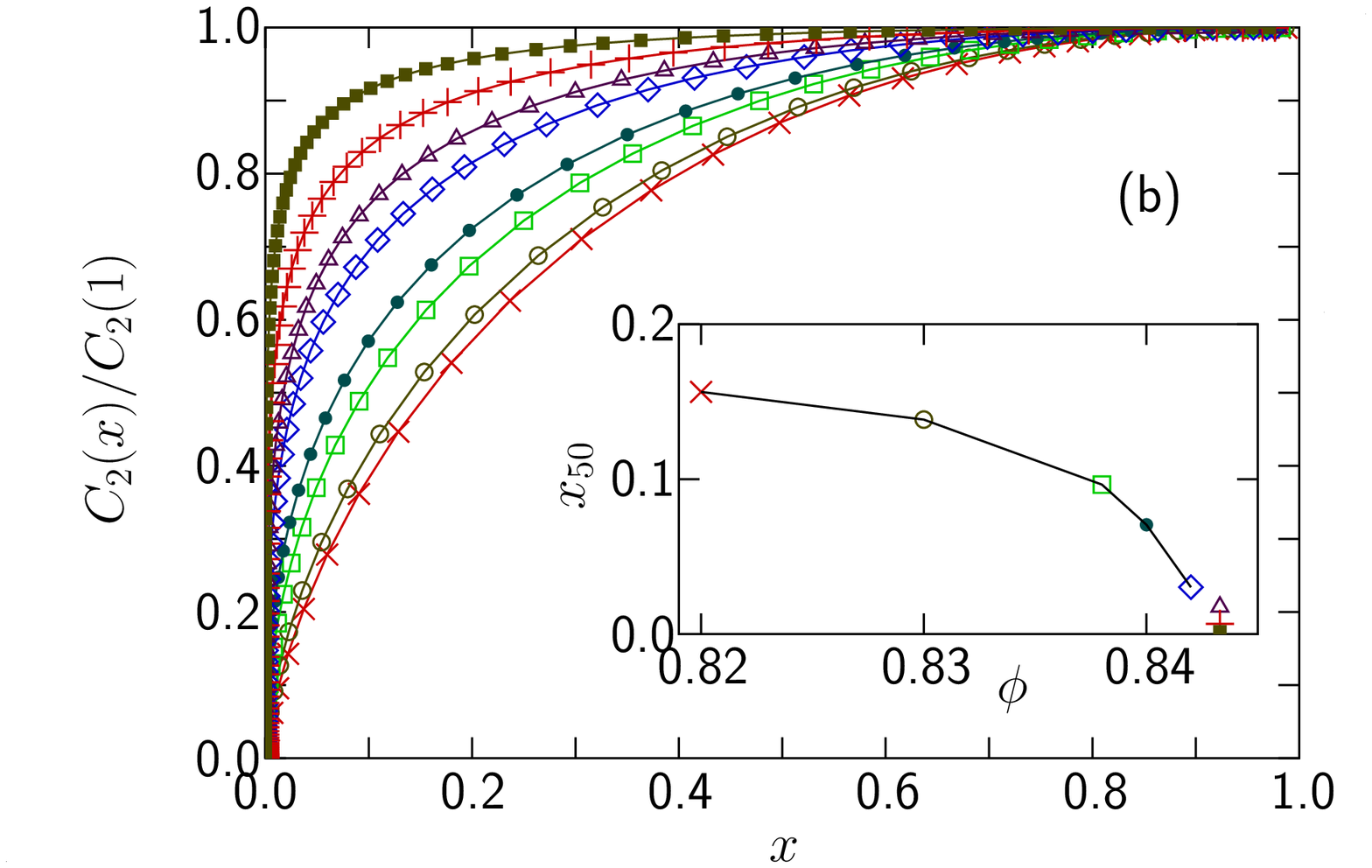}
  \includegraphics[bb=40 324 532 654, width=7cm]{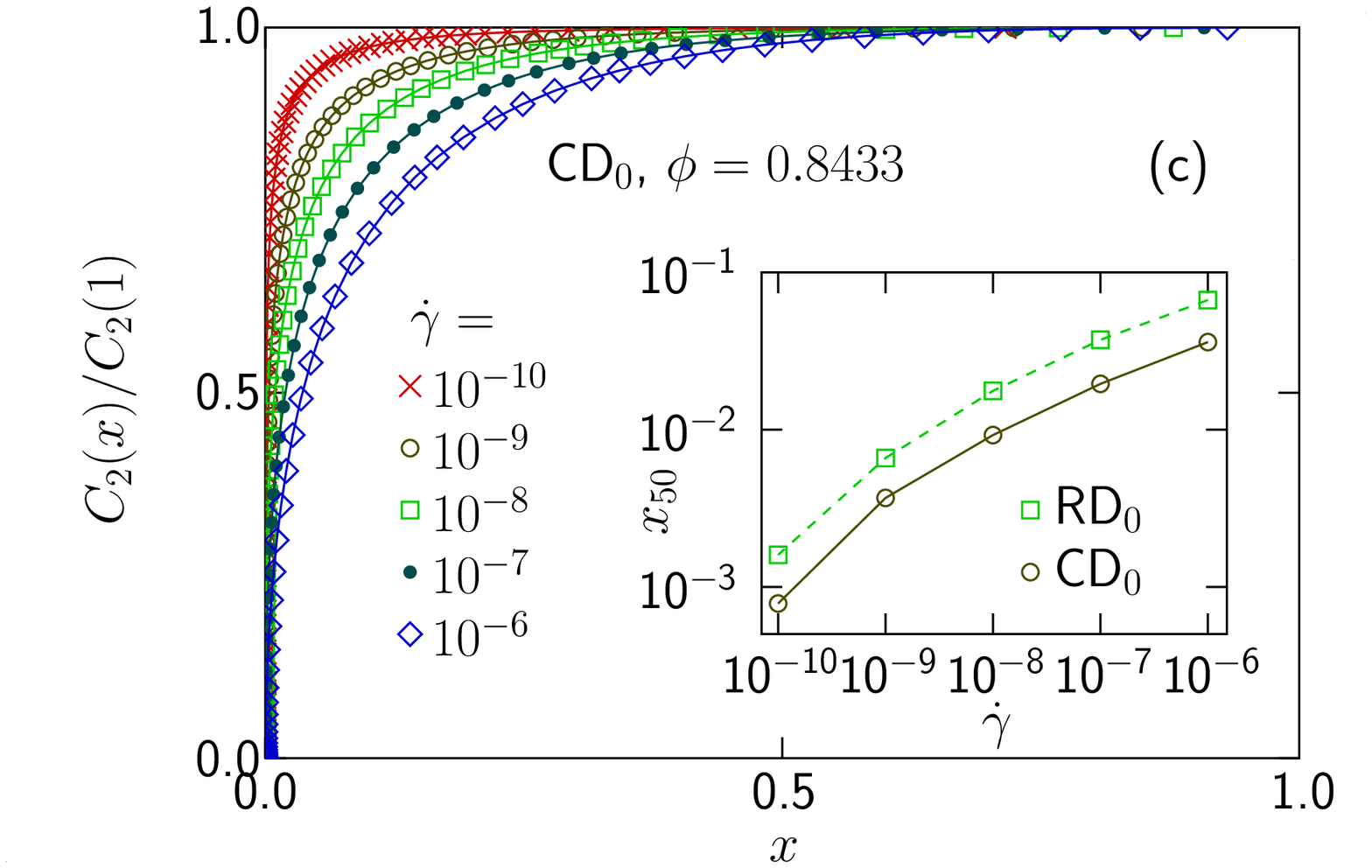}
  \caption{(Color online) Velocity distribution and dissipation. Panel (a) shows the
    velocity distribution function $P_\v(v)$ vs $v/\gdot$ with simulation parameters
    $(\phi,\gdot)$ and symbols as shown by the inset. The dashed line has slope
    $-3$. Panel (b) shows the part of the dissipated power which is dissipated by the
    fraction $x$ of the fastest particles. Panel (c) shows the same quantity for the
    CD$_0$ model. These data are at $\phi=0.8433\approx\phi_J$ and five different shear
    rates. The insets of panels (b) and (c) are $x_{50}$---the fraction of particles
    needed to dissipate 50\% of the power. The inset of panel (b) shows that $x_{50}$ for
    the RD$_0$ model decreases as $\phi$ increases whereas the inset of panel (c) shows
    $x_{50}$ at $\phi_J$ decreasing with $\gdot$ for both RD$_0$ and CD$_0$.}
  \label{fig:v2-x}
\end{figure}

A key quantity in the present letter is the energy dissipation. We here just remark that
this is a central quantity due to the relation between dissipation and rheology from power
balance, $V \sigma\gdot=k_d \expt{\sum_i \v_i^2}$\cite{Ono_Tewari_Langer_Liu}, and we
therefore believe that the considerations here may be instrumental in developing a better
understanding of shear-driven jamming.

Our first key result is that most of the energy is dissipated by a small fraction of fast
particles and, furthermore, that the fraction of particles needed to dissipate a given
part of the power decreases as jamming is approached.  Note that ``fast'' is here used in
a relative sense. For low $\gdot$ all particles are slow, it is only $v/\gdot$ that can be
big.  To study the dissipation we introduce the velocity distribution function $P_\v(v)$
such that $P_\v(v)dv$ is the fraction of particles with velocity
$v\leq|\v|<v+dv$. \Fig{v2-x}(a) shows $P_\v(v)$ vs $v/\gdot$ both at five densities below
$\phi_J$, and for three different shear rates at $\phi_J$. (To get histograms of good
quality down to small $P_\v$ we use bins that are equally spaced in $\ln v$.) The
different simulation parameters $(\phi,\gdot)$ and their corresponding symbols are shown
in the inset of panel (a). The points connected by solid lines and dash lines,
respectively, show two different ways to approach jamming. The solid line connects
$(\phi,\gdot)$ at $\phi<\phi_J$ and at sufficiently low $\gdot$ to be very close
to the hard disk limit. The dashed line connects three points at
$\phi=0.8433\approx\phi_J$. Here jamming is approached as $\gdot\to0$.

To study the dissipation with focus on the fast particles we define
\begin{equation}
  \label{eq:v2}
  x(v) = \int_v^\infty P_\v(v')\,dv',\quad
  \bar{C}_2(v) = \int_v^\infty P_\v(v') v'^2\,dv'.
\end{equation}
Here $x(v)$ is the fraction of fast particles with $|\v|>v$ and $k_d\bar{C}_2(v)$ is
the dissipating power due to the same particles. We also define $C_2(x)
=\bar{C}_2(v(x))$, where $v(x)$ is the inverse of $x(v)$. \Fig{v2-x}(b) shows the
normalized $C_2$ vs $x$ for the data in panel (a). The faster particles always dominate
the dissipation but this effect becomes more pronounced---the curves get steeper---as
jamming is approached; a smaller fraction of particles is then needed for a given part of
the dissipation.  As a simple quantitative measure we introduce $x_{50}$, shown in the
inset of panel (b), as the fraction of the fastest particles that dissipates 50\% of the
power. For the hard disk limit (solid line) $x_{50}$ decreases as $\phi$ increases towards
$\phi_J$. The behavior of $x_{50}$ at $\phi=0.8433\approx\phi_J$ is shown by the open
squares in the inset of panel (c); $x_{50}$ decreases with decreasing $\gdot$ and gets as
low as $0.16$\% at the lowest shear rate, $\gdot=10^{-10}$.  We believe that this
localization of the dissipation to a few faster particles is related to plastic events or
avalanches that are found above $\phi_J$, as already speculated by
others\cite{Heussinger_Berthier_Barrat:2010}.

Panel (c) shows that the CD$_0$ model behaves similarly. In this model it is the velocity
differences of contacting particles that is the quantity of interest rather than the
non-affine velocity, and $C_2$ is defined analogously.  The main data in panel (c) is
$C_2(x)$ at $\phi_J$ for the CD$_0$ model which is very similar to the three data sets at
$\phi_J$ in panel (a). As a more detailed comparison the inset of panel (c) shows $x_{50}$
against $\gdot$ at $\phi_J$ for both the RD$_0$ model and the CD$_0$ model, and it is
clear that this fraction decreases with decreasing $\gdot$ in both models. The effect
studied here is thus not just peculiar to the simpler RD$_0$ model.\cite{CD-3D}

The evidence from \Fig{v2-x} strongly suggests that $C_2(x)/C_2(1)$ approaches a step
function as $\phi\to\phi_J$ and $\gdot\to0$, and this is the main result from the first
part of this Letter.
For $C_2(x)/C_2(1)$ to approach a step function the limiting distribution has to
have a tail
\begin{equation}
  \label{eq:Pjamming}
  P_\v(v) \sim v^{-3},
\end{equation}
since that would make $C_2(v)$ diverge. We note that experiments on dense granular flows
have led to similar conclusions\cite{Moka_Nott}.  Before turning to more elaborate
analyses we note that the dashed line in \Fig{v2-x}(a) with slope $=-3$ gives some support
for \Eq{Pjamming} as the limiting behavior at $\phi_J$ as $\gdot\to0$.

For the further analysis it is important to understand the origin of the wide
distribution. We note that the non-affine velocity in the RD$_0$ model is related to the
sum of all (repulsive) contact forces that act on the particle. The non-affine velocity of
particle $i$ is $\v_i= \sum_j \f^\mathrm{el}_{ij}/k_d$. Close to jamming, the forces on
most particles almost cancel one another out, and the total force is typically very small
compared to the average force, $f_i^\mathrm{el}\ll f^\mathrm{el}_{ij}$, as has also been
noted by others\cite{DeGiuli:2015}. There are however some particles for which the forces
don't balance one another out, and the velocity of these particles can then be much larger
than the average velocity.  The wide distribution is thus due to the big difference
between the individual forces and the typical total force.

A consequence of this picture is that the maximum velocity is bounded by the typical
$f_{ij}^\mathrm{el}$ which means that the possibly algebraic distribution is cut off by an
exponential factor $e^{-v/v_c}$, where $v_c\sim f_{ij}/k_d \sim p/(k_dd_s)$. (This also
suggests $v_c/\gdot\sim\eta_p$.) This behavior is seen in \Fig{v2-x}(a) as the
approximately rectilinear (i.e.\ algebraic) behaviors for intermediate values of $P_\v(v)$
turn into more rapid decays at higher velocities.  One therefore expects the tails in the
distributions to be described by $P(v) \sim v^{-r} e^{-v/v_c}$, and this exponential decay
becomes a complicating factor, when one attempts to determine $r$ from $P(v)$.

\begin{figure}
  \includegraphics[bb=56 320 320 580, width=4.2cm]{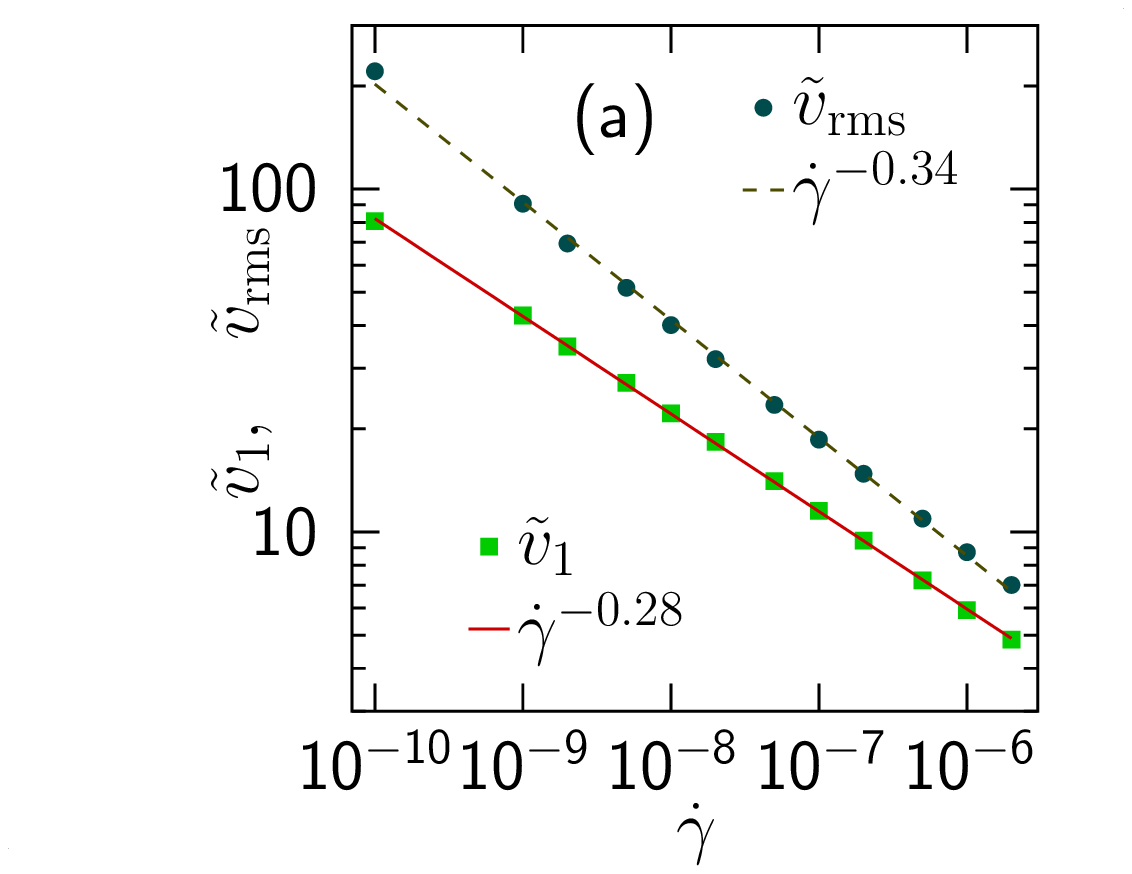}
  \includegraphics[bb=56 320 320 580, width=4.2cm]{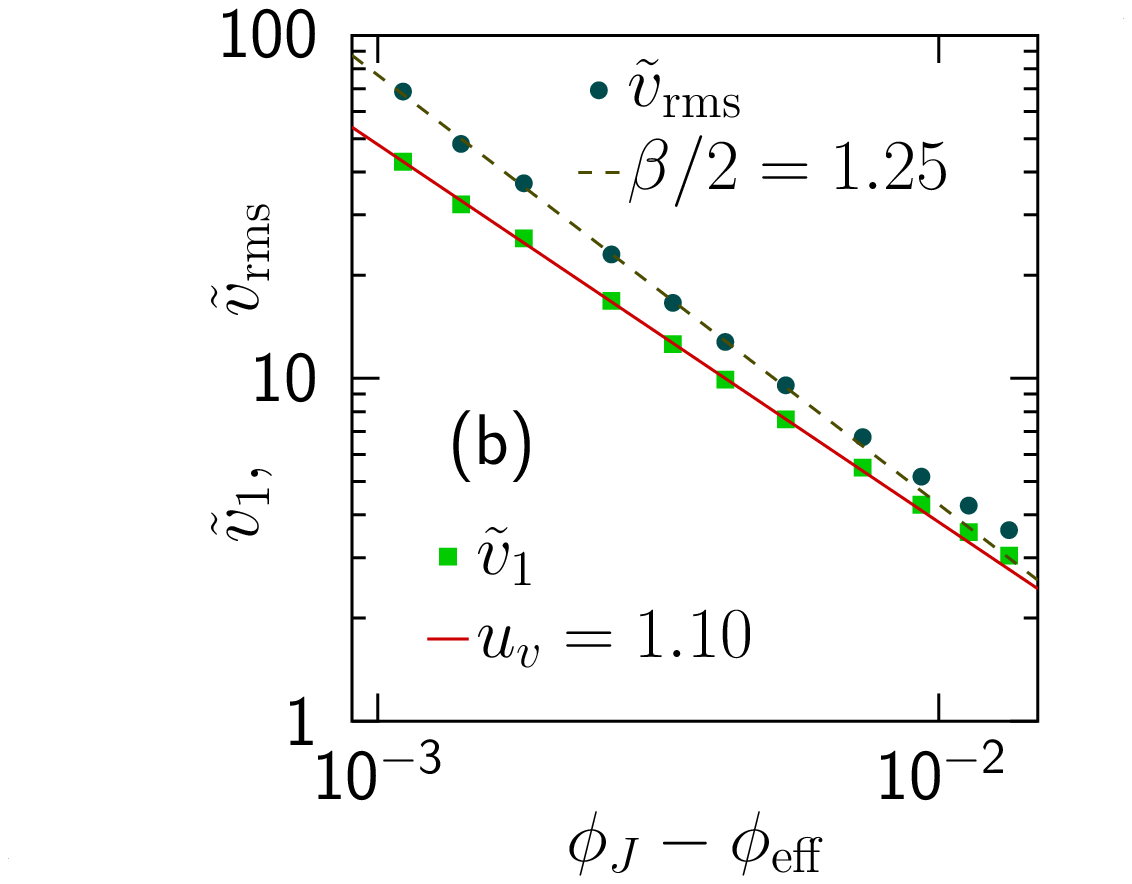}
  \includegraphics[bb=56 322 532 575, width=7cm]{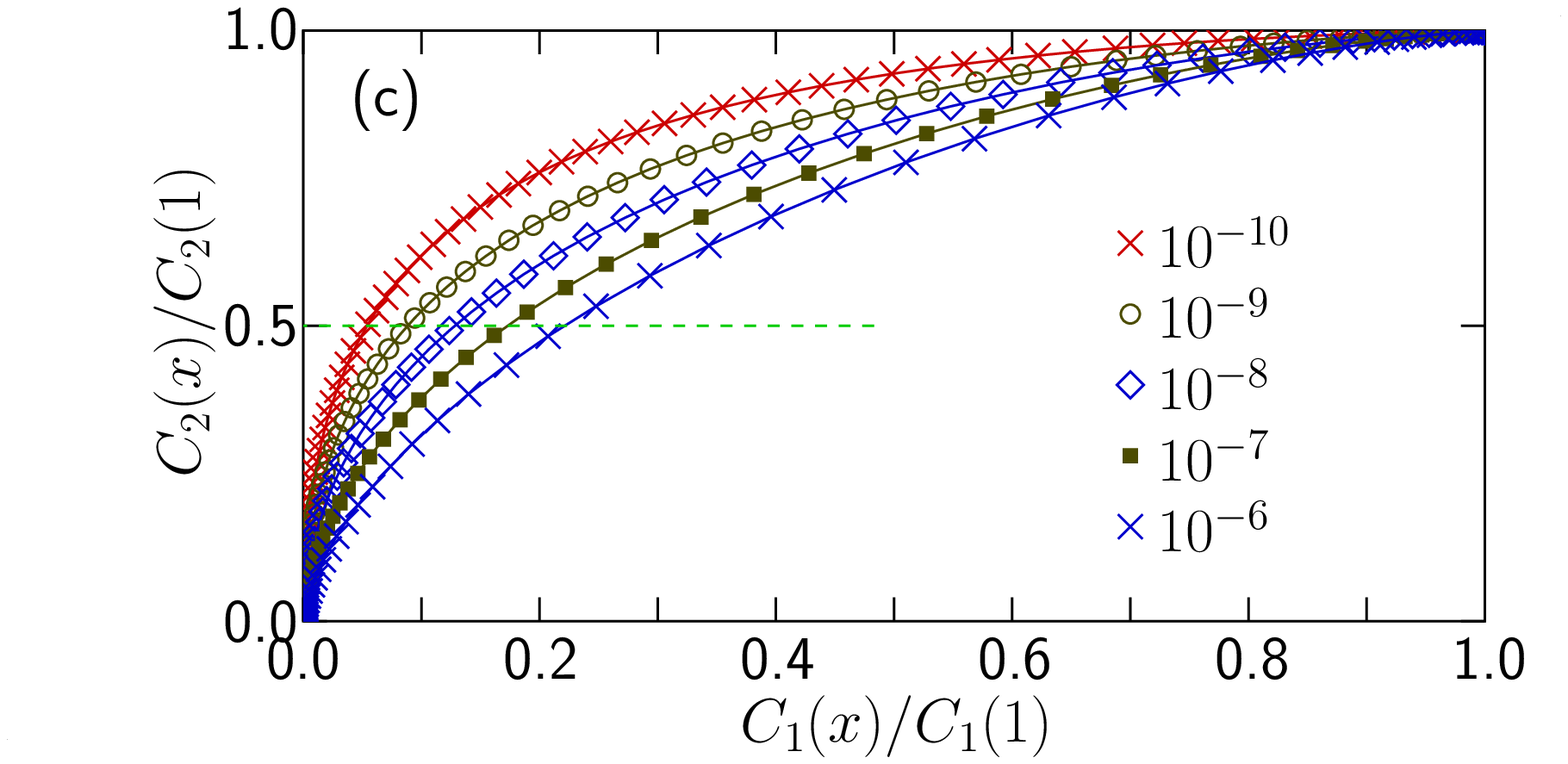}
  \caption{(Color online) Two different measures of the velocity. Panel (a) shows $\tvrms$
    and $\tvone$ for $\phi=0.8433\approx\phi_J$ vs $\gdot$. Panel (b) shows the same
    quantities for data below $\phi_J$ plotted vs distance to jamming; only the points
    with $\phi_J-\phieff<0.006$ were used for the fits. $\phieff$ is the effective
    density. The lines are $\sim(\phi_J-\phieff)^{-u_v}$ and
    $\sim(\phi_J-\phieff)^{-\beta/2}$. Panel (c) connects back to \Fig{v2-x}(a) but is
    $C_2$, related to $\tvrms^2$, against $C_1$ which is related to $\tvone$. From the
    crossings of the dashed line one may read off the relative contribution to $\tvone$
    from the fraction of the fastest particles that disspate 50\% of the energy---a
    quantity that decreases with decreasing $\gdot$.}
  \label{fig:v-compare}
\end{figure}

Our second key result is that different measures of the velocity behave differently. This
is important since it means that concepts like ``typical velocity''---used in various
theoretical approaches---then become useless.  
\Fig{v-compare} shows simulation results for $\tvone$ and $\tvrms$, defined through
\begin{equation}
  \label{eq:vy1}
  \tvone = \expt{|\tv|},\quad \mbox{and}\quad  \tvrms^2 = \expt{\tv^2},
\end{equation}
with the notation $\tv=\mathbf{v}/\gdot$. In panel (a) these quantities are plotted
against $\gdot$ and are found to diverge algebraically with different exponents:
$\tvrms\sim \gdot^{-\beta/2z\nu}\sim \gdot^{-0.34}$ and $\tvone\sim \gdot^{-u_v/z\nu}\sim
\gdot^{-0.28}$. [The expressions follow by taking $b=\gdot^{-1/z}$ in $\O(\phi, \gdot) =
b^{u_\O/\nu} g_\O((\phi_J-\phi) b^{1/\nu}, \gdot b^z)$\cite{Olsson_Teitel:gdot-scale} with
the scaling dimension $u_\O$ equal to $u_v$ for $\tvone$ and $\beta/2$ for $\tvrms$. The
latter follows from $\eta\sim\tvrms^2$ and $\eta(\phi,\gdot\to0)\sim
(\phi_J-\phi)^{-\beta}$.]

It is instructive to also examine the same quantities with data below $\phi_J$, close to
the hard disk limit. The starting point is the relations for hard disks,
$\tvone^\mathrm{hd}(\phi)\sim(\phi_J-\phi)^{-u_v}$, and
$\tvrms^\mathrm{hd}(\phi)\sim(\phi_J-\phi)^{-\beta/2}$,
that follow by using $b=\gdot$ in the scaling expressions and considering $\gdot\to0$.  In
\Fig{v-compare}(b) we make use of the effective-density mapping of soft disks onto hard
disks, $\O^\mathrm{hd}(\phieff) = \O(\phi,\gdot)$, where the effective density is $\phieff
= \phi-c E^{1/2y}$, with $c=1.53$ and $y=1.09$, as detailed in
Ref.~\cite{Olsson_Teitel:jam-HB}.  \Fig{v-compare}(b) shows $\tvone$ and $\tvrms$ against
$\phi_J-\phieff$. The solid line gives the exponent $u_v=1.10$ in agreement with
$\ell_\Delta \sim (\phi_J-\phi)^{-1.1}$ for the particle ``velocity'' in
Ref.~\cite{Heussinger_Berthier_Barrat:2010}. The dashed line gives $\beta/2=1.25$.  (The
value $\beta=2.50$, is somewhat low in comparison to recent
estimates\cite{Olsson_Teitel:gdot-scale}, but this could be due to not including
corrections to scaling\cite{Olsson_Teitel:gdot-scale}.)  Note that the exponents from
\Fig{v-compare}(a) and (b) are consistent when using
$1/z\nu=0.26$\cite{Olsson_Teitel:gdot-scale}.

The reason for the different behavior of $\tvone$ and $\tvrms$ is that the dominant
contribution to these quantities come from different velocity intervals. This is
illustrated in \Fig{v-compare}(c) which shows how $C_2(x)$ in \Eq{v2} and $C_1(x)$ (for
$v^1$ instead of $v^2$) increase to their respective limits $C_2(1)\equiv \tvrms^2$ and
$C_1(1)\equiv \tvone$, as $x$ (the fraction of particles included in the calculations)
increases.  The different curves get steeper for smaller $\gdot$ and for $\gdot=10^{-10}$
we find $C_1(x)\approx 0.05$ when $C_2(x)=0.5$ which thus shows that $\tvrms^2$ gets a
considerably larger contribution from the highest velocity part of the histogram than
$\tvone$. An extrapolation of these curves to the $\gdot\to0$ limit would give a step
function (though this is not as clearly suggested by the data as in \Fig{v2-x}) which
would imply that $\tvrms^2$ and $\tvone$ were controlled by different velocity intervals,
and that there is no reason for these quantities to be at all related.

\begin{figure}
  \includegraphics[bb=40 324 532 654, width=7cm]{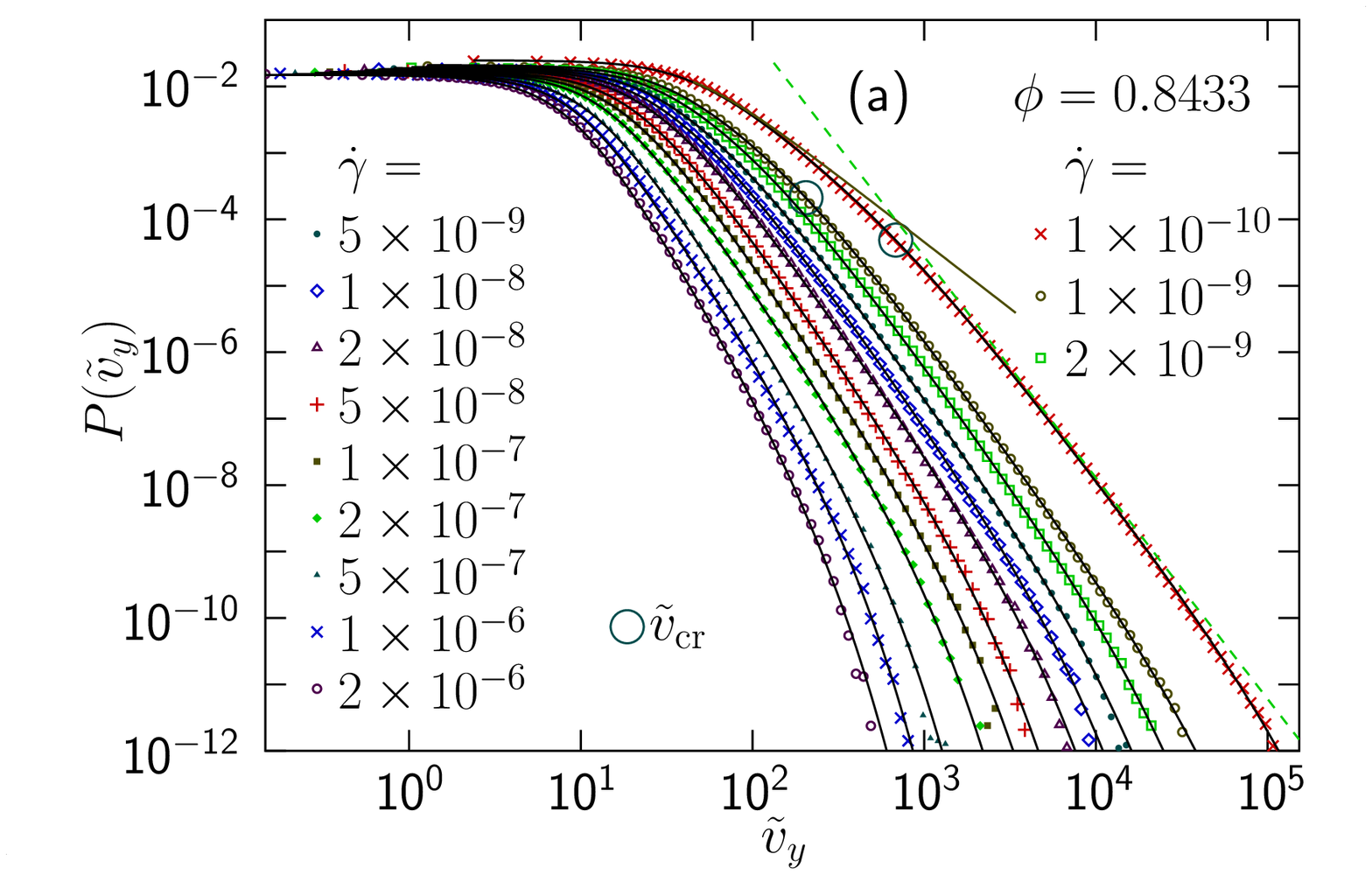}
  \includegraphics[bb=30 326 350 654, width=4.2cm]{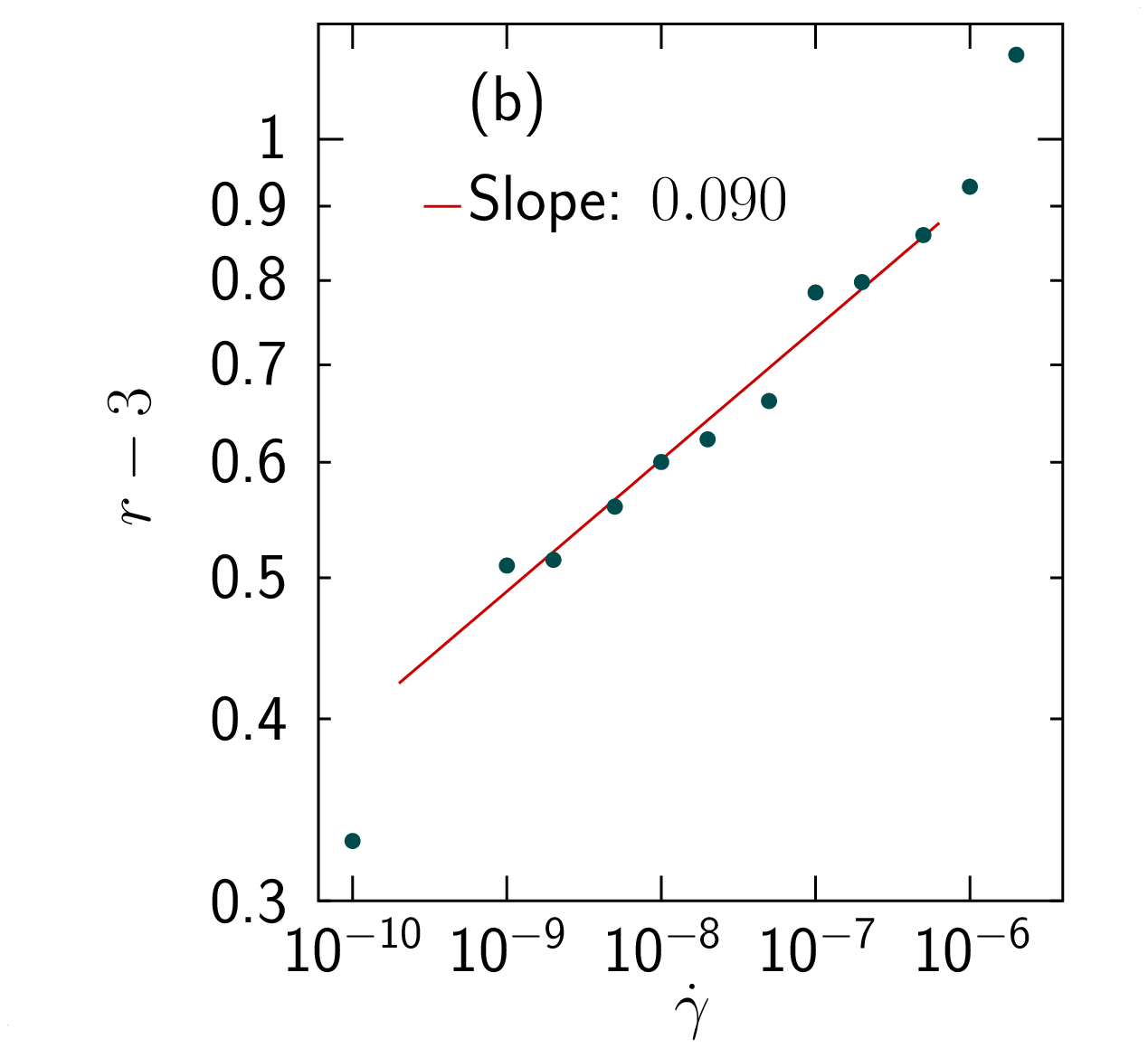}
  \includegraphics[bb=30 326 350 654, width=4.2cm]{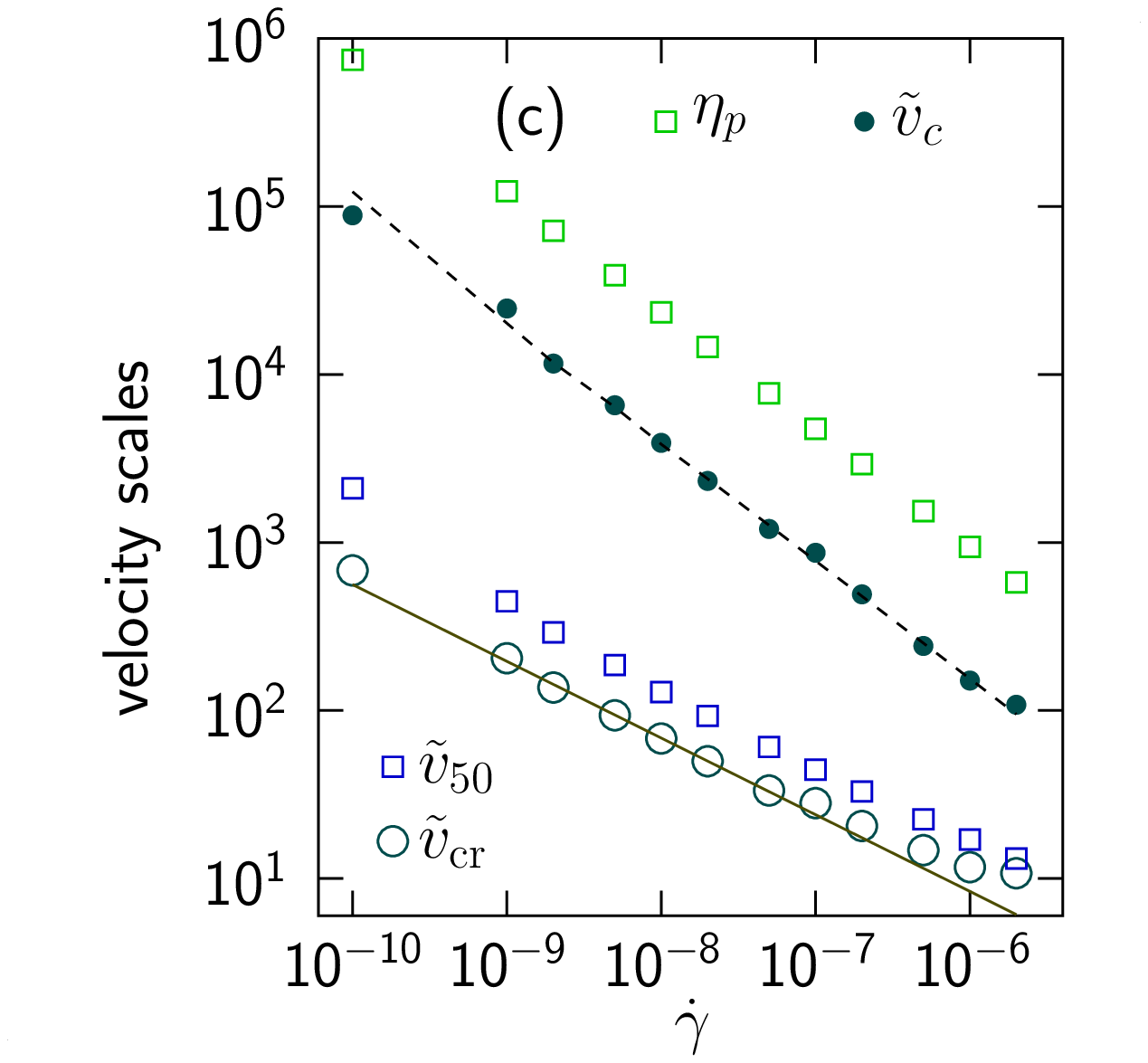}
  \caption{(Color online) Velocity distribution and the determination of the exponent $r$
    for different $\gdot$ at $\phi=0.8433\approx\phi_J$. Panel (a) is the velocity
    distribution $P(\tvy)$. Panel (b) shows $\tvc$ and $r-3$ vs $\gdot$ from fitting to
    \Eq{Afit}. Fitting $r-3\sim\gdot^{q_r}$ for $10^{-9}\leq\gdot\leq 5\times10^{-7}$
    gives $q_r=0.09\pm0.02$. Panel (c) shows that $\eta_p$ and $\tvc$ are
    proportional---the dashed line through $\tvc$ is $\eta_p/6.1$. Also shown are
    $\tvcr$---which is where the histogram crosses over to exponent $-r$, and $\tvhalf$
    above which 50\% of the dissipation takes place. These similarity of these two
    quantities suggests that the dominant part of the dissipation is given by particles
    governed by the algebraic tail.}
  \label{fig:vhist-fit}
\end{figure}

We will now relate our two key results of \Figs{v2-x} and \ref{fig:v-compare} to
properties of the velocity distribution function with the goal (1) to examine how the
exponent in \Eq{Pjamming} approaches $-3$ as $\gdot\to0$ (this exponent will be denoted by
$-r$) and (2) to shed some more light on the mechanism that allows $\tvone$ and $\tvrms$
to diverge differently. We have then found it convenient to use $P(\tvy)$---the
distribution of the absolute value of the $y$-component. This quantity differs from $P_\v$
in that it approaches a constant at small velocities---a feature that makes it easier to
find an analytical expression that fits the data.  \Fig{vhist-fit}(a) shows $P(\tvy)$ for
several different $\gdot$ at $\phi=0.8433\approx \phi_J$ together with the solid lines
that are fits to the expression
\begin{equation}
  \label{eq:Afit}
  P(\tvy) = \frac{A\; e^{-\tvy/\tvc}}{1 + (\tvy/\tva)^2 + (\tvy/\tvs)^r},
\end{equation}
with $A$, $\tvc$, $\tva$, $\tvs$, and $r$ as free parameters. This expression crosses over
from a constant at small $\tvy$ to a large-$\tvy$ tail with $\tvy^{-r} e^{-\tvy/\tvc}$ (as
discussed above) and the crossover is governed by an additional term in the denominator,
$(\tvy/\tva)^a$. For best possible fits, $a$ should be an additional free parameter, but
since $a$ anyway tends to be close to 2 and $a=2$ opens up for analytical
calculations\cite{Supp}, we here fix $a=2$.

\Fig{vhist-fit}(b) shows the exponent as $r-3$ vs $\gdot$.  The rectilinear behavior
suggests an algebraic decay, $r-3\sim\gdot^{q_r}$ with $q_r=0.09\pm0.02$, consistent with
the limiting behavior of \Eq{Pjamming}.  The points on top of \Fig{vhist-fit}(c) are the
cutoff velocity, $\tvc$ (solid circles) from the fits and $\eta_p$ (squares) directly from
simulations. The dashed line through the solid circles, which is $\eta_p/6.1$ and not a
fit to the data, confirms the expectation that these quantities should behave the
same. The covariation of $r$ and $\tvc$ (compare panels (b) and (c)) makes the fitting
difficult---a small decrease in $r$ can be compensated by a small decrease in $\tvc$ since
a smaller $r$ gives a slower decay, while a smaller $\tvc$ gives a faster decay. This
effect is most problematic at the lowest shear rate, $\gdot=10^{-10}$, and this point is
therefore not included in the determination of $q_r$.

It is now interesting to determine the size of the region governed by the algebraic decay,
$\tvy^{-r}$, and we therefore calculate the crossover velocity $\tvcr =
(\tvs^r/\tva^2)^{1/(r-2)}$, that describes the crossover from exponent $-2$ to exponent
$-r$, by equating the two velocity-dependent terms in the denominator of
\Eq{Afit}. \Fig{vhist-fit}(c) shows that $\tvcr$ behaves about the same as
$\tvhalf$---related to $x_{50}$ above--- which is the velocity above which 50\% of the
dissipation takes place. We thus find that the dissipation is largely governed by the
particles in the algebraic tail.  Recalling the conclusions from \Fig{v2-x}, it is clear
that the fraction of particles in the algebraic tail decreases with $\gdot$ and this is
also shown by the big open circles in \Fig{vhist-fit}(a) which are $P(\tvcr)$ vs $\tvcr$
for $\gdot=10^{-9}$ and $10^{-10}$---the last two points in a persistent trend to smaller
$P(\tvcr)$. The fact that the fraction of particles in the algebraic tail decreases with
decreasing $\gdot$ means that their contribution to $\tvone$ decreases ($\tvone$ is
instead dominated by the slower particles) whereas they always give the dominant
contribution to $\tvrms^2$\cite{Supp}. This explains the different behavior of $\tvone$
and $\tvrms$.  We finally note that the algebraic tail actually becomes wider as $\gdot$
decreases. \Fig{vhist-fit}(c) shows that $\tvc$ increases faster than $\tvcr$ which means
that the width of the algebraic tail---the region between $\tvcr$ and $\tvc$--- increases
with decreasing $\gdot$.

To summarize, we have found that the fraction of particles that are responsible for the
energy dissipation decreases towards zero as jamming is approached. These particles belong
to a tail in the velocity distribution that approaches $P(v)\sim v^{-3}$ at jamming. We
further find that different measures of the velocity diverge differently which means that
concepts like ``typical velocity'' no longer appear to be useful---a result of importance
for analytical approaches to shear-driven jamming.

\begin{acknowledgments}
  I thank S. Teitel for many illuminating discussions.  This work was supported by the
  Swedish Research Council Grant No.\ 2010-3725. Simulations were performed on resources
  provided by the Swedish National Infrastructure for Computing (SNIC) at PDC and HPC2N.
\end{acknowledgments}

\bibliography{j,this}
\bibliographystyle{apsrev4-1}

\end{document}


\section*{Supplemental material}
\begin{quote}
  Derivations related to ``Dissipation and velocity distribution at the shear-driven
  jamming transition'', Peter Olsson.
\end{quote}

\subsection*{Summary}

\begin{figure}[h]
  \includegraphics[height=5cm]{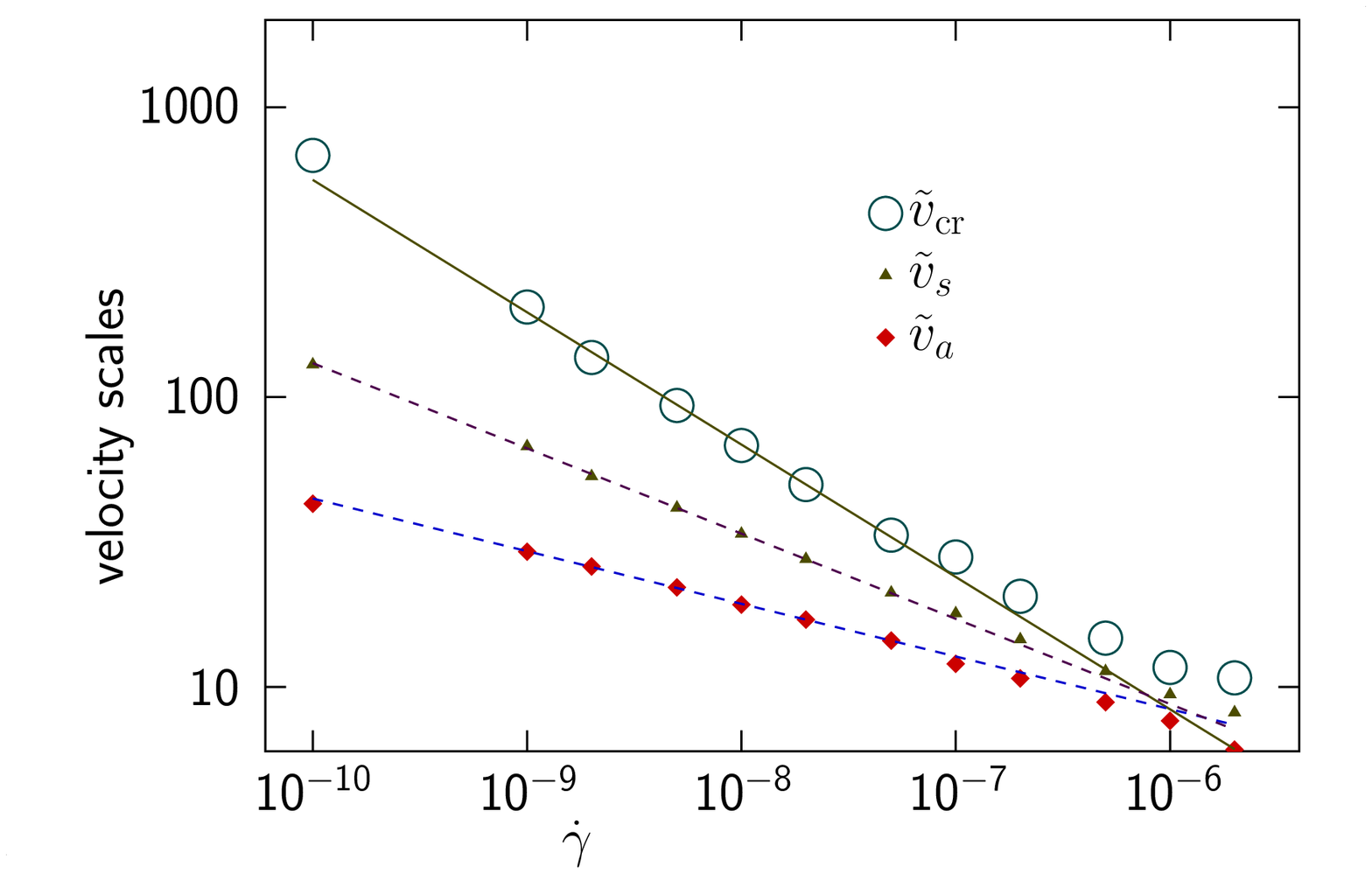}
  \caption{Velocity scales $\va$, $\vs$, and $\vcr$ of Eq.~(6). All these quantities
    behave algebraically for the smaller $\gdot$, to decent approximations.}
  \label{fig:vscales}
\end{figure}

The purpose of the calculations below is to illustrate the mechanism that gives different
behaviors for $\tvone$ and $\tvrms$. The idea is here to split the expression for the
velocity distribution
\begin{displaymath}
  P(\vv) = \frac{e^{-\vv/\vc}}{1+ (\vv/\va)^2 + (\vv/\vs)^r},
\end{displaymath}
into three different parts
\begin{equation}
  \label{eq:P-simp-a2}
  P(\vv) = \left\{
    \begin{array}{ll}
      1/[1+ (\vv/\va)^2], &\quad 0\leq\vv<\vcr,\\
      1/(\vv/\vs)^r, &\quad \vcr<\vv<\vc,\\
      0, &\quad \vc<\vv.
    \end{array}
    \right.
\end{equation}
We may then determine $\tvone=\expt{\vv}$ and $\tvrms^2=\expt{\vv^2}$ by calculating some
integrals analytically, as shown below. The results are then written in terms of the
different velocity scales, $\va$, $\vs$, $\vcr$, and $\vc$. \Fig{vscales} and Fig.~3(c)
show that all these quantities to decent approximations diverge algebraically for small
$\gdot$. We have $\va\sim\gdot^{-q_a}$, $\vs\sim\gdot^{-q_s}$,
$\vcr\sim\gdot^{-q_\mathrm{cr}}$, and $\vc\sim\gdot^{-q_c}$ with the exponents $q_a\approx
0.18$, $q_s\approx0.29$, $q_\mathrm{cr}\approx 0.46$, and $q_c\approx 0.71$.

The first moment of the velocity becomes
\begin{displaymath}
  \tilde v_1 = \expt \vv  \approx \va\left[ \ln\left(\frac{\vcr}{\va}\right) +
    \frac{1}{r-2}\right]
  \approx \va\left[0.62 \log_{10}(1/\gdot) + \frac{1}{r-2}\right].
\end{displaymath}

For the second moment of the velocity we get different results when $r$ is close to 3. For
$r$ not too close to 3 (which is the case relevant for comparisions with our simulation
data) we find
\begin{equation}
  \tvrms^2 \approx \va \vcr \left[1 + \frac{1}{r-3}\right],
\end{equation}
whereas for $r\to3$ we instead find
\begin{eqnarray}
  \tvrms^2 
  & \approx & \va \vcr + \frac{\vs^3}{\va}\ln\left(\frac{\vc}{\vcr}\right)
  \approx \va\vcr \left[1+(q_c-q_\mathrm{cr})\ln(1/\gdot)\right]\nonumber \\
  & \approx & \va\vcr \left[1+0.58\log_{10}(1/\gdot)\right].
\end{eqnarray}

\subsubsection*{The small-$\gdot$ limit}

With the values for the exponents given above, the leading small-$\gdot$ behaviors become
$\tvrms^2 \sim \va\vcr \sim \gdot^{-(q_\mathrm{cr}+q_a)/2} \sim \gdot^{-0.64}$ in very
good agreement with $\tvrms\sim \gdot^{-0.34}$ in Fig.~2(a).

For $\tvone$, the full expression $\tvone \sim \va [\ln(\vcr/\va)+1/(r-2)]$ reproduces the
exponent $-0.28$ in Fig.~2(a). In the $\gdot\to0$ limit, however, the second term vanishes
and the slowly changing logarithm may be replaced by a constant, and this gives $\tvone
\sim \gdot^{-q_a} \sim \gdot^{-0.18}$, which is clearly different from the measured
exponent, $-0.28$.  We consider the exponent obtained directly from the measured data to
be more reliable as it is from an excellent fit to data across four orders of magnitude in
$\gdot$.  In contrast, the ``analytical'' value for the exponent ($-0.18$) is obtained by
the dangerous process of extrapolating results from fitting to an expression (Eq.~(6))
that is used without any theoretical justification.

\newpage
\subsection*{Derivations}

To determine $\tvone\equiv\expt{\vv}$ and $\tvrms^2=\expt{\vv^2}$ we need the integrals
\begin{displaymath}
  I_0=\int P(\vv)\,d\vv,\quad I_1 = \int P(\vv) \vv\, d\vv,\quad I_2 = \int P(\vv) \vv^2\, d\vv,
\end{displaymath}
which with \Eq{P-simp-a2} becomes
\begin{displaymath}
  I_p = \int P(\vv)\vv^p d\vv = \int_0^{\vcr} \frac{\vv^p}{1+(\vv/\va)^2} d\vv +
  \int_{\vcr}^{\vc} \vv^p (\vv/\vs)^{-r} d\vv.
\end{displaymath}
It is then convenient to consider the two terms above separately.
\subsubsection*{First term, $I_p^{(1)}$}
We here use $x=\vv/\va$, $dx=d\vv/\va$, and $\xcr=\vcr/\va$ and handle the different
integrals separately for different $p$:
\begin{itemize}
\item With $p=0$:
\begin{eqnarray}
  I_0^{(1)} = \int_0^{\vcr} \frac{1}{1+(\vv/\va)^2} d\vv 
  & = & \va \int_0^{\xcr} \frac{1}{1+x^2} dx 
  = \va [\arctan x]_0^{\xcr} \nonumber \\
  & = & \va \arctan(\vcr/\va)
\end{eqnarray}
\item With $p=1$:
\begin{eqnarray}
  I_1^{(1)} = \int_0^{\vcr} \frac{\vv}{1+(\vv/\va)^2} d\vv 
  & = & \va^2 \int_0^{\xcr} \frac{x}{1+x^2} dx = \va^2 \left[\frac12
    \ln(1+x^2)\right]_0^{\xcr} \nonumber \\
  & = & \frac{\va^2}{2} \ln(1+(\vcr/\va)^2).
\end{eqnarray}
\item With $p=2$:
\begin{eqnarray}
  I_2^{(1)} = \int_0^{\vcr} \frac{\vv^2}{1+(\vv/\va)^2} d\vv 
  & = & \va^3 \int_0^{\xcr} \frac{x^2}{1+x^2} dx
  = \va^3 \left[x-\arctan x\right]_0^{\xcr} \nonumber \\
  & = & \va^2 \vcr - \va^3\arctan(\vcr/\va).
\end{eqnarray}
\end{itemize}

\subsubsection*{Second term, $I_p^{(2)}$}
We here get an expression for general $p$:
\begin{eqnarray*}
  I_p^{(2)} = \int_{\vcr}^{\vc} \vv^p (\vv/\vs)^{-r} d\vv 
  & = & \vs^r\left[\frac{\vv^{p+1-r}}{p+1-r}\right]_{\vcr}^{\vc}\\
  & = & \frac{\vs^r}{p+1-r}\left(\vc^{p+1-r}-\vcr^{p+1-r}\right)\\
  & = & \frac{\vs^{p+1}}{r-1-p} 
  \left[\left(\frac{\vs}{\vcr}\right)^{r-1-p}-\left(\frac{\vs}{\vc}\right)^{r-1-p}\right].
  \\
\end{eqnarray*}
This splits into two cases. If $r-1-p\gg0$ (which is always the case in our simulations)
and $\vc\gg\vcr$ (which allows us to skip the second term) we make use of
$\vcr^{r-2}=\vs^r/\va^2$ to get
\begin{equation}
  I_p^{(2)} \approx \frac{\vs^{p+1}}{r-1-p} \left(\frac{\vs}{\vcr}\right)^{r-1-p}
  = \frac{\va^2\vcr^{p-1}}{r-1-p}.
  \label{eq:I2_bigr}
\end{equation}
For $r-1-p\approx0$, on the other hand, (possible for $p=2$ and very close to jamming,
$r\to3$), we get
\begin{equation}
  I_p^{(2)} \approx \vs^{p+1} \left[\ln\left(\frac{\vs}{\vcr}\right) -
    \ln\left(\frac{\vs}{\vc}\right)\right]
  = \vs^{p+1} \ln\left(\frac{\vc}{\vcr}\right)
  \label{eq:I2_r3}
\end{equation}

\subsubsection*{Taking both terms together}

The normalization becomes
\begin{equation}
  I_0 = \va \arctan\left(\frac{\vcr}{\va}\right) + \frac{1}{r-1} \frac{\va^2}{\vcr}
  \approx \frac{\pi}{2}\va.
  \label{I0}
\end{equation}
For the first moment we get
\begin{eqnarray}
  I_1 & = & \frac{\va^2}{2} \ln\left[1+\left(\frac{\vcr}{\va}\right)^2\right] + \frac{\va^2}{r-2} \nonumber \\
  & \approx & \va^2\left[ \ln\left(\frac{\vcr}{\va}\right) + \frac{1}{r-2}\right].
  \label{I1}
\end{eqnarray}
The logarithmic correction is $\ln(\vcr/\va) = \ln(\gdot^{-q_\mathrm{cr}}/\gdot^{-q_a})
\sim (q_\mathrm{cr}-q_a)\ln(1/\gdot)$. The average velocity becomes (using $q_a\approx
0.18$ and $q_\mathrm{cr}\approx 0.46$)
\begin{displaymath}
  \tilde v_1 = \expt \vv  = I_1/I_0 \approx \va\left[ \ln\left(\frac{\vcr}{\va}\right) +
    \frac{1}{r-2}\right]
  \approx \va\left[0.62 \log_{10}(1/\gdot) + \frac{1}{r-2}\right].
\end{displaymath}

For the second moment and for $r$ not too close to 3, \Eq{I2_bigr} leads to
\begin{equation}
  I_2 = \va^2 \vcr \left[1 + \frac{1}{r-3}\right] - \va^3\arctan\left(\frac{\vcr}{\va}\right),
  \label{I2}
\end{equation}
and since $\arctan x<\pi/2$, the second term may be neglected and we get
\begin{equation}
  \expt{\vv^2} \equiv \tvrms^2 
  \approx \va \vcr \left[1 + \frac{1}{r-3}\right],
\end{equation}
For $r\to3$ we instead use \Eq{I2_r3} which gives (with $\vcr\approx \vs^3/\va^2$ and $q_c=0.71$).
\begin{eqnarray}
  \expt{\vv^2} \equiv \tvrms^2 
  & \approx & \va \vcr + \frac{\vs^3}{\va}\ln\left(\frac{\vc}{\vcr}\right)
  \approx \va\vcr \left[1+(q_c-q_\mathrm{cr})\ln(1/\gdot)\right]\nonumber \\
  & \approx & \va\vcr \left[1+0.58\log_{10}(1/\gdot)\right].
\end{eqnarray}